\documentclass[preprint,showpacs,aps]{revtex4}
\usepackage{amsmath,amssymb,epsfig}
\begin{document} 
\title[ Base  pair opening and bubble transport  in a DNA double  helix  induced by a protein molecular chain ]
{Base pair opening and bubble transport  in  a DNA double helix  induced by a protein molecule in a viscous medium}

\author{
V. Vasumathi}
\email{ vasu@cnld.bdu.ac.in}
\author{M. Daniel}
\email{daniel@cnld.bdu.ac.in}
\affiliation{ Centre for Nonlinear Dynamics, School of Physics, 
Bharathidasan University, Tiruchirappalli - 620 024, India.
}

\date{\today} 
\begin{abstract}
 We study the nonlinear dynamics of a protein-DNA molecular system   by treating  DNA as a set of two coupled linear chains and protein  in the form of a single linear chain  sliding along  the  DNA  at the physiological temperature in a viscous  medium. The nonlinear dynamics of the  above molecular system  in  general  is  governed by  a perturbed  nonlinear Schr\"{o}dinger equation. In the non-viscous limit, the equation reduces to the completely integrable nonlinear Schr\"{o}dinger (NLS) equation which admits N-soliton solutions.  The soliton excitations of  the DNA bases make localized base pair opening and travel along the DNA chain in the form of  a bubble. This may represent the bubble generated during the transcription process when an RNA-polymerase binds to a promoter site in the DNA double helical chain. The perturbed NLS equation is solved using a perturbation theory by treating the viscous effect due to surrounding  as a weak perturbation and the results show that the viscosity of the solvent in the surrounding damps out the amplitude of the soliton. 
\end{abstract}


\maketitle
\section{Introduction}
Protein-Deoxyribonucleic acid(DNA) interaction plays an important role in a large number of cellular processes such as
 gene expression, suppression, replication, transcription, recombination, repair and in several other 
 processes \cite{ref1}. DNA participation in the above processes are mediated or catalyzed by DNA-binding proteins
 like polymerases, helicases, nucleases, isomerases, ligases and histones.  Usually cellular
 processes   start with the  binding of a protein (enzyme) to the  DNA. For example, the transcription process  starts
 with the binding of RNA-polymerase (an enzyme protein) with a promoter site of the  DNA and this binding  is known to change the conformation of DNA by opening the bases. The above conformation change or base pair opening stresses the relevance of dynamics in understanding   protein-DNA interaction.
 Experimentally, these conformational changes in DNA during protein-DNA interaction have been studied through  atomic force microscopy \cite{ref2}, dynamic spectroscopy \cite{ref3}, X-ray crystallography of protein-DNA co-crystal \cite{ref4}, NMR  studies \cite{ref5} and electrophoresis  experiments \cite{ref6} as well as through fluorescent measurements \cite{ref6a}.   Results from free energy calculations combined with molecular dynamics simulations also explain the base flipping
 through protein-DNA interaction \cite{ref7,ref8,ref9,ref10}.  
However, from a theoretical point of view, the understanding of protein-DNA interaction and its dynamics is still  in its infancy. This is because the individual dynamics of DNA and   protein itself  is a very complex one. Also, interaction of the DNA molecule with the surrounding viscous medium and thermal fluctuations add to the complexity. The knowledge of the  DNA dynamics through soliton-like excitations  describe base pair opening or base flipping during  DNA functions. In this direction, Englander et al \cite{ref11},  used soliton excitations to explain base pair opening in DNA for the first time. Yomosa \cite{ref12,ref13} proposed a plane base rotator model by taking into account the rotational motion of bases in a plane normal to the helical axis, and  then Takeno and Homma \cite{ref14,ref15}   generalized the same  and the nonlinear molecular excitations were  shown to be governed by kink-antikink solitons. In contrast to the above, Peyrard and Bishop \cite{ref16} and Christiansen and his collegues \cite{ref17} proposed a model  by taking into account the transverse and longitudinal motions of bases in DNA to describe the base pair opening  through breather. Later, several authors\cite{ref18,ref20,ref21} including the present authors \cite{ref22,ref23}  suggested that either kink soliton or breather would be a good candidate to play a basic role in base pair opening in DNA.  In all the above studies, the  dynamics of DNA molecular system  has been studied  without  taking into account  thermal fluctuations and viscosity of the surrounding medium  and hence, it was shown that, the soliton representing the base pair opening in DNA may travel for infinite distance and time.   Recently, few authors studied the effect of thermal fluctuations \cite{cm,jm1} and viscosity of the surrounding medium   on  unzipping and soliton-like base-pair opening and showed that these effects damp the solitons and hence travel  only for a limited distance \cite{jm2,ref19,sm1,sm2,vs1}.   Like DNA dynamics, the dynamics of protein through soliton excitations also plays an important role in describing energy transfer in alpha helix proteins.  In this regard, Davydov \cite{dvs,dvs1}  who for the first time used soliton  excitations to explain energy transfer in alpha helical proteins by proposing a  new model by  considering the coupling between the quantum transition occurring due to the vibrational structure of the $C=O$ bond and  the elastic longitudinal wave propagation along the chain with the helix behaving like a spring. Later on, the above study was extended by several authors \cite{dvs2,dvs3} including one of the present authors \cite{dvs4,dvs5} to  study the effect of exchange excitation between the chains,  temperature, higher order interactions and interspine coupling. In contrast to Davydov's model, Yomosa \cite{toda1} and  very recently Sataric et al \cite{toda2} studied  soliton excitations in protein by considering  the molecule as a toda lattice chain.    The recent results on  the statistical   mechanics of protein-DNA system through coarse-grain and worm-like chain models describe  unzipping of the DNA  chain \cite{sp,sp1,sp2,sp3}. Normally, protein molecule interacts  with DNA either through  non-specific interaction (sequence-independent) which is   mainly driven by the electrostatic attractive force between positively charged amino acids and negatively charged phosphate groups of the DNA back bone or through specific interactions (sequence-dependent) including hydrogen bonds, van der Waals force and water mediated bonds between the  protein molecule and specific site of DNA \cite{muru}.   In a recent paper, Sataric et al \cite{ref24} studied the impact of regulatory proteins through hydrogen bonds on breather excitations in DNA by considering Davydov model of amide-I vibration for protein dynamics and Peyrard-Bishop's model of stretching of hydrogen bonds  for DNA dynamics and found that binding of protein to DNA gives rise to a large-amplitude breather.  Except the above, no author has so far studied the impact of protein on the dynamics of DNA. Hence, in the present paper, we   study the conformational changes  that take place in the form of base  pair opening in DNA through nonlinear excitations induced by a  protein molecule through interaction  at the  physiological temperature in a viscous  surrounding medium.  The paper is organized as follows. In section II, we  present details of the above  model and the Hamiltonian for the  DNA molecular chain. The model  Hamiltonian for protein-DNA interaction is given in section III.  In section IV, we derive the  equation of motion  for the protein-DNA molecular system  in the continuum limit and   in the next section (section V)   base pair opening  in DNA  as soliton solution of the  associated  nonlinear dynamical equation in a non-viscous medium is shown. The effect of viscosity of the surrounding medium on the base pair opening in DNA is understood through perturbation analysis in section VI.  The results are concluded in section VII.
\section{Model  Hamiltonian for DNA  dynamics}  
We consider  a DNA double helix  in B-form along with a protein molecule (say RNA-polymerase) at physiological temperature in a surrounding viscous medium and study the impact of protein interaction on the dynamics of DNA.  The model we propose here for the study treats  DNA as a  set of two coupled linear  molecular chains and  protein as a single linear chain  interacting   through a linear  coupling.  A schematic representation of the above protein-DNA molecular system  is shown in Fig. 1(a). In the figure  $R$ and $R'$  represent the two complementary strands in the DNA double helix. Each arrow  represents the direction of the base attached
     to the strands and the dots between arrows represent  the net hydrogen bonding effect between the 
     complementary bases.
 The shaded ellipse overlapping the DNA double helical structure  represents interaction of protein with the DNA molecule. 
 The conformation and stability of  DNA double helix is  mainly determined by the stacking
     of bases through  intrastrand dipole-dipole interaction and through  hydrogen bonds  between the  complementary bases (interstrand interaction). From a heuristic argument, it was assumed that the  hydrogen bonding energy between the complementary bases  depends on  the distance between them. Generally,  the distance between the complementary bases  can be expressed through  longitudinal,    transverse and rotational motion of bases. Among the different motions, the rotational motion of bases is found to contribute more towards the opening of bases pairs. Hence, for our study, we consider a plane-base rotator model for DNA \cite{ref12,ref14} which the authors have extensively used for studying pure DNA dynamics in the recent times \cite{ref22,ref23}. In order to find the distance between  the complementary bases, in Figs. 1(b) and 1(c), horizontal
     projections  of the $n^{th}$ base pair in the xy and xz-planes are presented respectively. In the figure,    $Q_n$ and  $Q'_n$  denote the  tip of the $n^{th}$ bases belonging to  the
     complementary strands $R$ and $R'$  at $P_n$ and $P'_{n}$ respectively.  Let $\theta_n(\theta'_n)$ and $\phi_n(\phi'_n)$ represent the
     angles of rotation of the bases in the  $n^{th}$ base pair  in the xz and xy-planes  respectively.
 \begin{figure}
\begin{center}
\epsfig{file=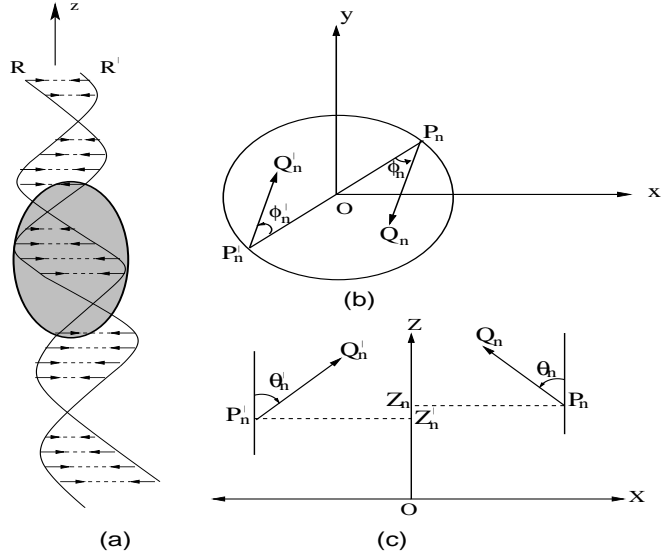,height =8cm, width=8.5cm}
\caption{(a) A schematic representation of a DNA double helix molecule  with a protein  molecule interacting with it. (b) A horizontal projection of the $n^{th}$ base
pair of the DNA in the xy-plane. (c) A horizontal projection of the $n^{th}$ base pair of DNA in the xz-plane.}
\end{center}
\end{figure} 
        By using   simple geometry in Figs. 1(b,c), we can write  the distance between the tips of bases as \cite{ref15}
\begin{eqnarray}
 ( Q_nQ'_n)^{2}&=&2+4 r^{2}+(z_n-z'_n)^2+2 (z_n-z'_n)\nonumber\\
&&\times\left(\cos\theta_n -\cos\theta'_n\right) -4r\left[\sin\theta_n\cos\phi_n\right.\nonumber\\
&&\left.+\sin\theta'_n\cos\phi'_n\right]+2\left[\sin\theta_n\sin\theta'_n
\left(\cos\phi_n\right.\right.\nonumber\\
&&\left.\times\cos\phi'_n+\sin\phi_n\sin\phi'_n\right)
\left.-\cos\theta_n\cos\theta'_n\right],\label{eq1a}
\end{eqnarray}
where `$r$' is the radius of the circle depicted in Fig. 1(b). \\

 The hydrogen bonding  energy can be understood in a more clear and  transparent way by introducing  quasi-spin operators  ${\bf
S_n}\equiv(S_n^{x}, S_n^{y}, S_n^{z})=(\sin\theta_n\cos\phi_n,~   \sin\theta_n\sin\phi_n, ~ \cos\theta_n)$ and  $ {\bf S'_n}\equiv(S_n^{'x}, S_n^{'y}, S_n^{'z})=(\sin\theta'_n\cos\phi'_n, ~ \sin\theta'_n\sin\phi'_n, ~ \cos\theta'_n)$. Using  the above, Eq.(\ref{eq1a}) can be rewritten as
\begin{eqnarray}
(Q_n Q'_n)^2&=&2+4r^2+2\left[S_n^{x} S_n^{'x}+S_n^{y} S_n^{'y}
 -S_n^{z} S_n^{'z}\right]\nonumber\\
&&-4r\left[S_n^{x}+S_n^{'x}\right].\label{eq1b}
\end{eqnarray}
While writing Eq.(\ref{eq1b}), we have  chosen $z_n=z'_n$. It is interesting to note that the form of $({Q_nQ'_n})^2$ given in Eq.(\ref{eq1b}) is the same as the Hamiltonian for  a generalized form of the Heisenberg spin model. Therefore, the intrastrand base-base interaction  in DNA also can be written using the same consideration. Also, it is  reasonable to think that if such a quasi-spin model  can be used in this problem, the double strand DNA and the rung-like base pairs can be conceived as a  coupled spin chain model or a spin ladder system.

With the above consideration, we are at liberty to use the following  Heisenberg model of the
Hamiltonian for a coupled spin chain model or 
 spin ladder system with  ferromagnetic-type
exchange interaction  between nearest neighbouring spins in the same lattice (equivalent to stacking of bases in one strand i.e. intrastrand interaction) and ferromagnetic or antiferromagnetic rung-coupling (equivalent to hydrogen bonds between  complementary bases i.e. interstrand interaction).

\begin{eqnarray}
H_D=-\sum_n \left[ J ({\bf S_n\cdot S_{n+1}+S'_n\cdot S'_{n+1}})+\mu ({\bf S_n\cdot S'_n})\right].\label{eq1}
\end{eqnarray}
 Thus, the DNA double helical chain is mapped onto a two coupled spin chain model or a spin ladder system with ferromagnetic legs ($J>0$) and ferromagnetic ($\mu>0$) or antiferromagnetic  ($\mu<0$) rungs. Therefore, in Hamiltonian (\ref{eq1}), the terms proportional to $J$ correspond to stacking interaction between the $n^{th}$ base and its nearest neighbours in the two strands and the last term which is proportional to $\mu$ corresponds to the interstrand interaction or hydrogen bond energy between the complementary bases.  In equilibrium, the parameter $\mu$ is expected to be less than zero (corresponding to antiferromagnetic coupling).\\
As DNA works at the biological temperature, the hydrogen atom attached to the bases are also normally in a thermally excited state. Therefore, it is necessary to generalize the above model into a thermal DNA. Thus, to include the effect of thermal phonons into the system, we add the following Hamiltonian.
\begin{subequations}
\begin{eqnarray}
H_T&=&\sum_n \left[\frac{p_n^2}{2m_1} +k_1(X_n-X_{n+1})^2\right],\label{eq2a}\\
H_{D-T}&=&\alpha_1 \sum_n  (X_{n+1}-X_{n-1})({\bf S_n\cdot S'_n}),\label{eq2b}
\end{eqnarray}
\end{subequations}
where $p_n=m _1\dot X_n$, with overdot representing the time derivative and $k_1$ represents the elastic constant. $m_1$ is the mass of the  hydrogen atom attached to the base  and $X_n$ represents the displacement of it  at the $n^{th}$ site  along the direction of  the hydrogen bond. The interaction Hamiltonian  $H_{D-T}$ given in Eq. (\ref{eq2b})  represents the coupling between the vibration of the above hydrogen atom (thermal fluctuation) and the rotation of bases.
\section{Model  Hamiltonian for Protein-DNA interaction}
 As protein binding to DNA induces large mechanical stress on  it, which  causes conformational changes, and thus  acts as a precursor for all the functions of DNA. Hence, we investigate  here the dynamics of a DNA when a  protein molecule  interacts with it  by sliding on the DNA chain. Eventhough,  proteins are  much larger  in size than the DNA(substrate), only a small portion of the protein molecule, namely the active site is directly interacting with the DNA molecule (see Fig.1 (a)).  Hence, we consider the short active site region of the protein molecule  that interacts with the DNA as a linear chain as has been treated by several other authors (see for e.g. Sataric et al \cite{toda2}). Thus, as said earlier, we propose  the  model for the above protein-DNA  molecular system  by considering DNA as a set of two coupled linear chains and the active site region of the protein  molecule as a linear  molecular chain interacting  with the bases  which is  schematically shown in Fig. 2. The DNA part of the sketch is the linearized chains of Fig. 1(a) and $n, n\pm1$ represent the $n^{th}$ and $(n\pm1)^{th}$ sites representing the bases (similar to $P_n, P'_n, P_{n\pm1}, P'_{m\pm1}$ in Figs. 1(b,c) ).   $G$ represents the linear protein  molecular chain (active site region)   interacting with the bases by sliding on the  DNA chains.  To explain  the model for the above system further, we consider the protein molecule as a collection of mass points, with each mass point representing a peptide unit and connected by linear springs exhibiting longitudinal stretching motion parallel to the helical axis of DNA, i.e. along z-direction which couples to  the hydrogen bonds of bases in a linear way.
    Hence,  the model Hamiltonians for the longitudinal stretching motion of the   protein molecule ($H_p$) and its interaction with the DNA chain ($H_{D-p}$) is written as
\begin{subequations}
\begin{eqnarray}
H_p&=&\sum_n \left[\frac{q_n^2}{2m_2} +k_2(y_n-y_{n+1})^2\right],\label{eq3a}\\
H_{D-p}&=&\alpha_2 \sum_n  (y_{n+1}-y_{n-1})S_n^z S_n^{'z},\label{eq3b}
\end{eqnarray}
\end{subequations}
where $q_n=m _2\dot y_n$, and  $m_2$ is the mass of the peptide.   $y_n$ denotes  displacement of the $n^{th}$ peptide in the protein chain from its equilibrium position   and $k_2$ represents the elastic constant  associated with the small amplitude oscillation of the protein molecule.
\begin{figure}
\begin{center}
\epsfig{file=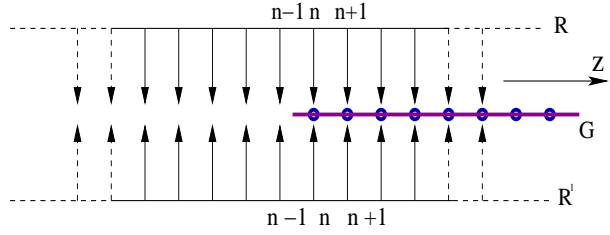,height =3.0cm, width=8cm}
\caption{A sketch representing the protein-DNA molecular system}
\end{center}
\end{figure} 
The interaction Hamiltonian  $H_{D-p}$ given in Eq.(\ref{eq3b}) is chosen to represent the change in hydrogen bonding energy due to change in the distance between the adjacent peptide units along the hydrogen bonding spines of the protein molecule and $\alpha_2$ is the coupling coefficient. Further, as the protein molecule is assumed to slide on the DNA chain, the interaction  energy along z-direction is expected to be dominant over the energy in the xy-plane normal to it (see Eq.(\ref{eq3b})). 
 Thus,  the total Hamiltonian for   our model   can be written using   Eqs. (\ref{eq1}), (\ref{eq2a}), (\ref{eq2b}), ({\ref{eq3a}}) and (\ref{eq3b}) as
\begin{eqnarray}
H&=& H_D+H_p+H_{D-p}+H_T+H_{D-T},\nonumber\\
&=&\sum_n \left[ -\{J ({\bf S_n\cdot S_{n+1}+S'_n\cdot S'_{n+1}})+\mu ({\bf S_n\cdot S'_n})\}\right. \nonumber\\
&&+\frac{p_n^2}{2m_1}+\frac{q_n^2}{2m_2}+k_1 (X_n-X_{n+1})^2+k_2 (y_n-y_{n+1})^2\nonumber\\
&&\left.+\alpha_1 (X_{n+1}-X_{n-1})({\bf S_n\cdot S'_n}) +\alpha_2   (y_{n+1}-y_{n-1})S_n^z S_n^{'z}\right].\label{eq4}
\end{eqnarray}
~~\\
Before proceeding further, for the sake of completeness, we present the form of the Hamiltonian (\ref{eq4}) in terms of angles of rotation of bases as
\begin{eqnarray}
H&=&\sum_n\left[-J \{ \sin\theta_n\sin\theta_{n+1} \cos
(\phi_{n+1}-\phi_n)\right.
 + \cos\theta_n\cos\theta_{n+1}\nonumber\\
&&+\sin\theta'_n\sin\theta'_{n+1}  
\cos (\phi'_{n+1}-\phi'_n)+ \cos\theta'_n\cos\theta'_{n+1}\}\nonumber\\
&&-[\mu-\alpha_1  (X_{n+1}-X_{n-1})] \{\sin\theta_n\sin\theta'_n 
\cos (\phi_n-\phi'_n)+\cos\theta_n\cos\theta'_n\}\nonumber\\
&& +\frac{p_n^2}{2m_1}+\frac{q_n^2}{2m_2}+k_1 (X_n-X_{n+1})^2+k_2 (y_n-y_{n+1})^2\nonumber\\
&&\left.+\alpha_2  (y_{n+1}-y_{n-1})\cos\theta_n\cos\theta'_n\right],\label{eq4a}
\end{eqnarray}
In the case of Heisenberg spin systems, when the spin value is large, the spin dynamics is understood either through a classical approach or under semi-classical approximation by bosonizing the Hamiltonian (see for e.g. Ref. \cite{mdlk}).  Also, it should be mentioned that creation and annihilation operators were used  to represent  the Hamiltonian  while studying the transport of charge and hole along short DNA molecules \cite{hol1,hol2} and while investigating the nonlinear dynamics of alpha helical protein molecules using the model proposed by Davydov \cite{dvs}. Therefore, along the same lines,
in order to understand the dynamics of the above protein-DNA molecular system, here also, we bosonize the Hamiltonian (\ref{eq4}) using  Holstein-Primakoff (H-P) representation \cite{hp} for quasi spin operators by writing $
{S}_{n}^{+}=\sqrt{2}\,[1-\epsilon^{2}a_{n}^{\dagger}a_{n}]^{1/2}\epsilon a_{n},
{S}_{n}^{-}=\sqrt{2}\,\epsilon a_{n}^{\dagger}[1-\epsilon^{2}a_{n}^{\dagger}a_{n}]^{1/2},  
 {S}_{n}^{z}=[1-\epsilon^{2}a_{n}^{\dagger}a_{n}],
$
where ${S}_{n}^{\pm}={S}_{n}^{x}\pm{i{S}_{n}^{y}} $. In the low temperature limit, $a_n^{\dagger}a_n<< 2S,$ and hence, the H-P transformation can be expanded in a power series  in terms of the parameter $\epsilon=1/\sqrt{S}$ as
\begin{subequations}
\begin{eqnarray}
 S^{+}_n &=&\sqrt{2}\epsilon[1-\frac{\epsilon^2}{4}a_n^{\dagger} a_n-\frac{\epsilon^4}{32}a_n^{\dagger}a_n a_n^{\dagger} a_n-O(\epsilon^6)] a_n, \label{eq5a}\\
 S^{-}_n &=&\sqrt{2}\epsilon a_n^{\dagger}[1-\frac{\epsilon^2}{4}a_n^{\dagger} a_n-\frac{\epsilon^4}{32}a_n^{\dagger}a_n a_n^{\dagger} a_n-O(\epsilon^6)],\label{eq5b}
\end{eqnarray}
\end{subequations}
 and similar expansions  for $S'^{+}_n, S'^{-}_n$ and $S'^{z}_n$ in terms of $b_n (b_n^{\dagger})$. Here $a_n^{\dagger}(b_n^{\dagger})$ and $a_n(b_n)$ represent creation and annihilation operators of the $n^{th}$ bases and  satisfy the usual commutation relations, $[a_m,a_n^{\dagger}]=[b_m,b_n^{\dagger}]=\delta_{mn}, [a_m,a_n]=[b_m,b_n]=[a_m^{\dagger},a_n^{\dagger}]=[b_m^{\dagger}, b_n^{\dagger}]=0.$ 
Substituting  Eqs. (\ref{eq5a}) and (\ref{eq5b}) in Eq.(\ref{eq4})   , we have the Hamiltonian  upto   O$(\epsilon^2)$ as
\begin{eqnarray}
H&=&\sum_n \left[\frac{p_n^2}{2m_1}+\frac{q_n^2}{2m_2} +k_1 (X_n-X_{n+1})^2+k_2 (y_n-y_{n+1})^2\right.\nonumber\\
&&+\epsilon^2\{-J(a_n a_{n+1}^{\dagger}
+a_n^{\dagger}a_{n+1}-a_n^{\dagger}a_n-a_{n+1}^{\dagger}a_{n+1}\nonumber\\
&& +b_n b_{n+1}^{\dagger}+b_n^{\dagger}b_{n+1}-b_n^{\dagger}b_n-b_{n+1}^{\dagger}b_{n+1})\nonumber\\
&&-[\mu-\alpha_1 (X_{n+1}-X_{n-1})](a_nb_n^{\dagger}+a_n^{\dagger}b_n-a_n^{\dagger}a_n\nonumber\\
&&\left.-b_n^{\dagger}b_n)-\alpha_2(y_{n+1}-y_{n-1}) (a_n^{\dagger}a_n+b_n^{\dagger}b_n)\}\right].\label{eq7}
\end{eqnarray}

\section{The dynamical equations}
 Having  written down the  Hamiltonian  in the semi-classical description, the dynamics of the protein-DNA  molecular system  can be understood by constructing the equations of motion as  
\begin{eqnarray}
i\hbar\frac{\partial a_{n}}{\partial t}&=&\Big[a_{n},H\Big],\nonumber\\
	&=& F\Big(a_{n}^{\dagger},a_{n},a_{n+1}^{\dagger},a_{n+1}\Big),\label{eq8a}
\end{eqnarray}
 and similar one for $b_{n}$. The equations of motion  for $X_n$  and $y_n$ are written using  the Hamilton's equations of motion $\frac{\partial X_n}{\partial t}=-\frac{\partial H}{\partial p_n},~ \frac{\partial p_n}{\partial t}=\frac{\partial H}{\partial X_n},~\frac{\partial y_n}{\partial t}=-\frac{\partial H}{\partial q_n} $ and $ \frac{\partial q_n}{\partial t}=\frac{\partial H}{\partial y_n} $.
The  explicit  form of the equations of motion  can be derived by substituting  Hamiltonian (\ref{eq7}) in the   above equations of motion  for $a_{n}, b_{n}, X_n $ and $y_n$.  Thus, we get
\begin{subequations}
\begin{eqnarray}
i \frac{\partial a_n}{\partial t} &=& - J(a_{n+1}-2 a_n+a_{n-1})-[\mu+\alpha_1 (X_{n+1}\nonumber\\
&&  -X_{n-1})](b_n-a_n)-\alpha_2 (y_{n+1}-y_{n-1}) a_n,\label{eq9a}\\
i \frac{\partial b_n}{\partial t} &=&-J (b_{n+1}-2b_n+b_{n-1})-[\mu+\alpha_1 (X_{n+1} \nonumber\\
&& -X_{n-1})](a_n-b_n)-\alpha_2 (y_{n+1}-y_{n-1}) b_n,\label{eq9b}\\
m_1\frac{\partial^2 X_n}{\partial t^2} &=& k_1 (X_{n+1}-2X_n+X_{n-1})+\alpha_1 [a^{\dagger}_{n-1} a_{n-1}\nonumber\\
&&-a^{\dagger}_{n+1}a_{n+1}+b^{\dagger}_{n-1} b_{n-1}-b^{\dagger}_{n+1}b_{n+1}\nonumber\\
&& a_{n+1}b_{n+1}^{\dagger}-a_{n-1}b_{n-1}^{\dagger}+a_{n+1}^{\dagger}b_{n+1}-a_{n-1}^{\dagger}b_{n-1}],\label{eq9c}\\
 m_2\frac{\partial^2 y_n}{\partial t^2} &=& k_2 (y_{n+1}-2y_n+y_{n-1})+\alpha_2 [a^{\dagger}_{n-1} a_{n-1}
\nonumber\\
&&-a^{\dagger}_{n+1}a_{n+1}+b^{\dagger}_{n-1} b_{n-1}-b^{\dagger}_{n+1}b_{n+1}].\label{eq9d}
\end{eqnarray}
\end{subequations}
While writing the above equations (\ref{eq9a}-\ref{eq9d}), we have rescaled the time variable  and redefined  $m_1, m_2,k_1$ and $k_2$. In order to represent the  large amplitude collective modes  by coherent states, 
 we  introduce  Glauber's coherent state representation  \cite{glab}  for boson operators $a_n^{\dagger}|u>=u_n^{*}|u>, a_n|u>=u_n|u>,|u>=\Pi_n|u_n> $ and $b_n^{\dagger}|v>=v_n^{*}|v>, b_n|v>=v_n|v>,|v>=\Pi_n|v_n> $ with $<u|u>=1$ and $<v|v>=1$ where $u_n$ and $v_n$ are the coherent amplitudes of the operators $a_n$ and $b_n$ for the system in the states $|u>$ and $|v>$ respectively. Further, as the length of the DNA and the protein chains  are very large compared to the lattice parameter, we make a continuum approximation by introducing the new fields  $u_n\rightarrow u(z,t), v_n( t)\rightarrow v(z, t), X_n (t)\rightarrow X(z,t )$ and $y_n( t)\rightarrow y(z, t)$, where $z=n l$  with the expansions $u_{n\pm 1}=u(z,t)\pm l \frac{\partial u}{\partial z}+\frac{l^2}{2!}\frac{\partial^2 u}{\partial z^2 }\pm O(l^3)$ and similar ones for $v_{n\pm 1},~X_{n\pm 1}$ and $y_{n\pm 1}$. Under the above approximations, the equations of motion (\ref{eq9a}), (\ref{eq9b}), (\ref{eq9c}) and (\ref{eq9d})  after rescaling $z$ and redefining $\alpha_1,\alpha_2,k_1$ and $k_2$ upto $O(l^2)$ can be written as
\begin{subequations}
\begin{eqnarray}
i u_{ t}&=&- u_{ z z}-(\mu-\alpha_1 X_z)(v-u)-\alpha_2 ~y_{ z} u, \label{eq10a}\\
i v_{ t}&=&- v_{ z z}-(\mu-\alpha_1 X_z)(u-v)-\alpha_2 ~y_{ z} v, \label{eq10b}\\
m_1 X_{ t t}&=& k_1 X_{ z z}-\alpha_1 ~[|u|^2+|v|^2-uv^{*}-u^{*} v]_{ z},\label{eq10c}\\
m_2 y_{ t t}&=& k_2 y_{ z z}-\alpha_2 ~[|u|^2+|v|^2]_{ z}.\label{eq10d}
\end{eqnarray}
\end{subequations}
In Eqs. (\ref{eq10a}-\ref{eq10d}), the suffices $t$ and $z$  represent partial derivatives with respect 
to time $t$
and the spatial variable $z$ respectively.
 On  subtracting Eq. (\ref{eq10a}) from (\ref{eq10b}) and by choosing  $v=-u$, Eqs. 12 (a-d) can be written as
\begin{subequations}
\begin{eqnarray}
i u_{ t}-\{2\mu -(\alpha_2 y_z-2\alpha_1 X_{ z})\}u+ u_{ z z} =0,\label{eq11a}\\
X_{ t t}-\frac{ k_1}{m_1} X_{ z z}=-\frac{4\alpha_1}{m_1} [|u|^2]_{ z},\label{eq11b}\\
y_{ t t}-\frac{ k_2}{m_2} y_{ z z}=-\frac{2\alpha_2}{m_2} [|u|^2]_{ z}.\label{eq11c}
\end{eqnarray}
\end{subequations}
 It may  be mentioned that, addition of Eqs. (\ref{eq10a}) and (\ref{eq10b}) satisfy identically.
The term proportional to $\mu$ in Eq. (\ref{eq11a}) can be transformed away using the transformation $u( z, t)=\hat u( z,  t)e^{-2 i\mu  t}$
   and the result reads (after dropping the hat)
\begin{eqnarray}
i u_{ t}+ u_{ z z}+(2\alpha_1 X_{ z}+\alpha_2 y_{z}) u =0,\label{eq12a}
\end{eqnarray}
The set of coupled equations (\ref{eq11b},c) and (\ref{eq12a})  describe the dynamics of our protein-DNA molecular system at the biological temperature,  when the protein molecule binds to the DNA double helical chain through linear harmonic coupling.
 The dynamics is found to be governed by   the excitation of DNA bases and thermal vibration of the hydrogen atoms attached to the bases   combined with the  longitudinal motion of  peptide units of the binding protein.
In particular, we are concerned with the nonlinear excitation of bases induced by protein and thermal fluctuations,  in which a cluster of DNA bases may undergo a large excursion as compared to the rest of the bases. It may be noted that when  $\alpha_1=\alpha_2=0$, Eqs.~(\ref{eq12a}), (\ref{eq11b}) and (\ref{eq11c})  are decoupled and  reduced to a set of linear equations.  Thus, when the protein molecule is detached from the DNA chain  ($\alpha_1=\alpha_2 =0$), the dynamics is governed by the following  well known set of linear equations.
\begin{subequations}
\begin{eqnarray}
i u_{ t}+ u_{ z z} &=&0,\label{eq13a}\\
X_{ t t}-\frac{ k_1}{m_1} X_{ z z}=0, ~y_{ t t}-\frac{ k_2}{m_2} y_{ z z}&=&0.\label{eq13b}
\end{eqnarray}
\end{subequations}
While Eq. (\ref{eq13a}) is the time-dependent  Schr\"{o}dinger  equation for  a free particle,  Eqs.(\ref{eq13b})  are the  homogeneous linear wave equations.
 Eq. (\ref{eq13a}) admits plane transverse  wave solution of the form $u=u_0 e^{i(\kappa z-w t)}$  with the dispersion relation $w=\kappa^2 $ where $u_0$ is the constant amplitude. On the other hand,   Eq. (\ref{eq13b}) admits linear non-dispersive wave solutions  $X= f_1( z-v_1 t)+g_1 ( z+v_1 t)$ and $y=f_2(z-v_2 t)+g_2(z+v_2 t) $ where $f_1, g_1$ and $f_2,g_2$ are arbitrary functions and $v_1=\sqrt{\frac{ k_1}{m_1}}$ and $v_2=\sqrt{\frac{k_2}{m_2}}$,  represent the constant phase velocities of the wave.
When the protein molecule started interacting with the  DNA molecular chain at the physiological temperature, i.e when $\alpha_1\neq 0 $ and $\alpha_2\neq 0$, the excitation energy of the protein-DNA molecular system  at the physiological temperature (thermal fluctuation) increases, and nonlinearity started playing its role. Thus, the set of full coupled nonlinear equations become important and it is essential that Eqs. (\ref{eq11b},c)  and (\ref{eq12a}) should be solved in their full form to  understand the underlying nonlinear dynamics.
\section{Soliton, Base Pair Opening and Bubble Transport} 
In order to solve the set of coupled equations (\ref{eq12a}), (\ref{eq11b}) and  (\ref{eq11c}), in their full form  we  differentiate Eqs. (\ref{eq11b}) and (\ref{eq11c}) with respect to  $z$  once and define  $\hat X(z,t)=X_z$ and $Y(z,t)=y_z$,  so that Eqs.  (\ref{eq12a}), (\ref{eq11b}) and (\ref{eq11c}) are written as
\begin{subequations}
\begin{eqnarray}
i u_t +u_{zz}+(2\alpha_1 \hat X+\alpha_2 Y) u=0\label{eq15a},\\
\hat X_{tt}-v_1 ^2 \hat X_{zz}+\frac{4\alpha_1}{m_1}[|u|^2]_{zz}=0,\label{eq15b}\\
Y_{tt}-v_2 ^2 Y_{zz}+\frac{2\alpha_2}{m_2}[|u|^2]_{zz}=0.\label{eq15c}
\end{eqnarray}
\end{subequations}
 Now, we  rewrite Eqs. (\ref{eq15b}) and (\ref{eq15c}) by introducing  the wave variable $\zeta= z-v_3  t$ and  writing $\hat X(z,t)\rightarrow \hat X(\zeta), Y( z, t)\rightarrow Y(\zeta)$ . 
\begin{subequations}
\begin{eqnarray}
\hat X_{\zeta\zeta}- 2\beta_1[|u|^2]_{\zeta\zeta}=0,\label{eq16a}\\
y_{\zeta\zeta}- 2\beta_2[|u|^2]_{\zeta\zeta}=0,\label{eq16b}
\end{eqnarray}
\end{subequations}
where $\beta_1=\frac{2\alpha_1}{m_1( v_1^2- v_3^2)}$ and $\beta_2=\frac{ \alpha_2}{m_2( v_2^2- v_3^2)}$.  On integrating Eqs. (\ref{eq16a}) and (\ref{eq16b}) with respect to $\zeta$ twice and assuming both  the integration constants to be zero, we get $\hat X=2\beta_1 |u|^2$ and $Y= 2\beta_2|u|^2$,  which upon using in Eq. (\ref{eq15a}) gives
\begin{eqnarray}
i  U_{t}+U_{zz} +2 |U|^2 U  =0.\label{eq17}
\end{eqnarray}
While writing  Eq. (\ref{eq17}),  we have made use of the transformation  $u(z,t)= (2\alpha_1\beta_1+\alpha_2\beta_2)^{-\frac{1}{2}}U(z,t)$.
\begin{figure}
\begin{center}
\epsfig{file=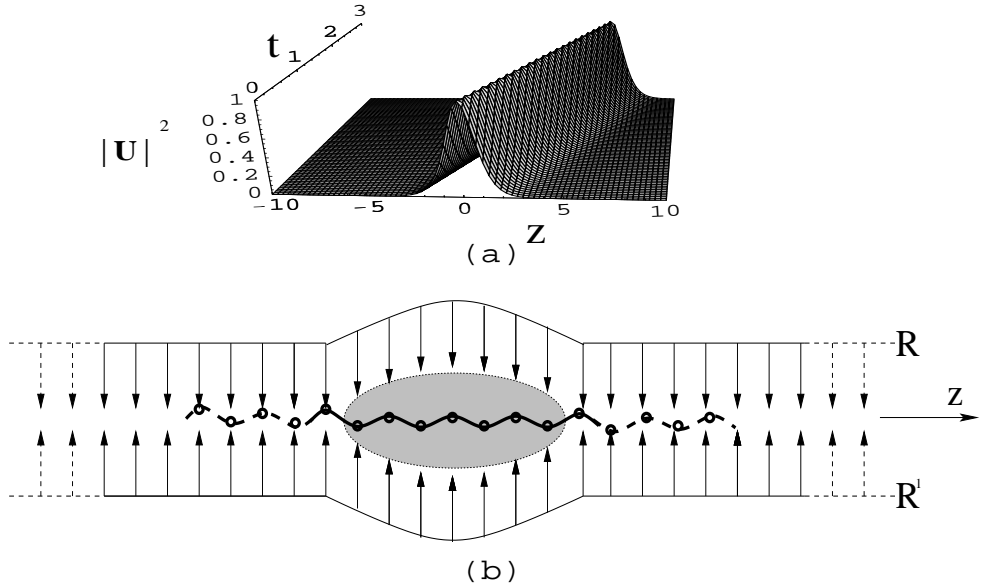,height =8cm, width=13cm}
\caption{ (a)  One soliton solution ( Eq. \ref{eq19}) of the NLS equation . (b) A schematic representation of  formation of  bubble  with the solitons and its propagation along DNA.}
\end{center}
\end{figure}
  Eq. (\ref{eq17}) is the well known  completely integrable nonlinear Schr\"{o}dinger (NLS) equation    which  has been  solved for N-soliton solutions using the inverse scattering transform (IST) method \cite{ref25}.    For instance, the  one soliton solution is written as 
\begin{eqnarray}
U=\eta sech[\eta( z-2\xi  t-\theta_0)]
 exp[i\xi( z-2\xi t-\theta_0)+i((\eta^2+\xi^2) t-\sigma_0)],\label{eq19}
\end{eqnarray}
where $\eta,\xi, \theta_0$ and $\sigma_0$ are  four real parameters which determine the propagating amplitude, velocity, initial position and initial phase of the soliton. The solitons in the protein-DNA molecular system with thermal fluctuation are formed  as a result of the dynamical balance between the dispersion due to interaction of intrastrand (stacking) dipole vibrations in each strand of the DNA with the nonlinearity  provided by the interaction between the hydrogen bonds in DNA and the local displacement of the peptide groups in the protein molecule and the thermal phonons. The longitudinal waves that arise in the protein molecule and hydrogen atoms in DNA in turn provide a potential well that prevents dispersion of the rotational energy of the bases in DNA. Thus, the propagation of rotation of bases in DNA is coupled to the longitudinal waves of protein and the coupled excitations propagate as a localized and dynamically self-sufficient entity called solitons which travel along  each strand of the DNA chain. In Fig. 3(a), we have plotted the square of the absolute value of the one soliton solution $U$ i.e. $|U|^2$   as given in Eq. (\ref{eq19}).  In Fig. 3(b), we present a schematic representation of the coherent base excitations in DNA in terms of rotation of bases induced by the protein molecule in the form of solitons propagating  along the two strands which collectively form a travelling bubble created by energy delocalization due to nonlinear effects.  Thus, the soliton solution describes an open state configuration in  the individual strands of the DNA double helix which collectively represent a bubble. In the figure, the shaded ellipse represents  the region of interaction of  the protein molecule  with DNA where the bubble is formed. Thus, the protein  molecule acts as a zip-runner  in opening the bases in  DNA chain   during the  process of transcription.  Similar results have been observed experimentally by Ha et al\cite{ref26}, on the  winding and unwinding of  E-coli Rep helicase-DNA complex. Further, our results on bubble propagation  in DNA due to protein interaction is in accordance with the experimental data   on the binding of RNA-polymerase to promoter \cite{ref27,ref28,ref29,ref30,ref31}. From the expression for the soliton namely $u= (2\alpha_1\beta_1+\alpha_2\beta_2)^{-\frac{1}{2}} \eta sech[\eta( z-2\xi  t-\theta_0)]
 exp[i\xi( z-2\xi t-\theta_0)+i((\eta^2+\xi^2) t-\sigma_0)]$, it is noted that the amplitude of the soliton depends on the coupling of the DNA excitations to the  thermal phonons ($\alpha_1$) and to the molecular vibrations of the protein ($\alpha_2$)  which were also respossible for nonlinearity in the soliton equation. For large coupling, it is expected that the amplitude of the soliton decreases. Recently Campa \cite{cm} also showed through simulation studies that, in the case of large thermal coupling, the bubble travels only for a short distance with decreasing amplitude.

\section{Effect of Viscosity }
In a more realistic description of the dynamics of  protein-DNA system, it is important to consider the effect of the surrounding medium or environment. Effectively, the interaction of DNA with the surrounding medium reduces to viscous damping effect. It is known that, in the case of a protein-DNA system the solvating water acts as a viscous medium that makes the nucleotide oscillations to damp out \cite{mp1}. The effect of viscous force exerted on the DNA chain can be taken into account by adding a term of the form $-i\gamma U$ to the right-hand side of the completely integrable nonlinear Schr\"{o}dinger equation (\ref{eq19})  which now becomes a damped nonlinear Schr\"{o}dinger equation   as given below.
\begin{eqnarray}
i U_t+U_{zz}+2|U|^2 U=-i\gamma U\label{eq20}
\end{eqnarray}
As the viscosity of water is temperature dependent, from a simple fluid mechanics argument,  one can estimate the magnitude of the damping coefficient as very small at the physiological temperature \cite{sm1}. Hence, we treat the term proportional to $\gamma$ in Eq. (\ref{eq20}) as a weak perturbation. When  $\gamma=0$  Eq. (\ref{eq20})  reduces to the completely integrable NLS equation as given in Eq. (\ref{eq17}) and the one soliton solution is given in Eq. (\ref{eq19}) which  can be rewritten  for convenience in the form
\begin{eqnarray}
U=\eta \mbox{sech}\eta( \theta-\theta_0)
 \exp[i\xi(\theta-\theta_0)+i(\sigma-\sigma_0)],\label{eq21}
\end{eqnarray}
where $\frac{\partial\theta}{\partial t}=-2\xi,~ \frac{\partial\theta}{\partial x}=1,~\frac{\partial\sigma}{\partial t}=\eta^2+\xi^2$ and $\frac{\partial\theta}{\partial x}=0$.\\
We
 carry out a perturbation analysis \cite{mdjb} to understand the impact of the viscous force by  introducing  a slow time variable $T=\gamma t$  and treat the quantities $\eta,\xi,\theta_0 $ and $\sigma_0$  as functions of this time scale and hence the envelope soliton solution  (\ref{eq21}) is written as
\begin{eqnarray}
U=\hat U (\theta,T;\gamma) \exp[i\xi(\theta-\theta_0)+i(\sigma-\sigma_0)].\label{eq22}
\end{eqnarray}
Under the above assumption of quasi-stationarity, Eq. (\ref{eq20}) reads
\begin{eqnarray}
\eta^2 \hat U+\hat U_{\theta\theta}+2|\hat U|^2 \hat U=\gamma F(\hat U),\label{eq23}
\end{eqnarray}
where
\begin{eqnarray}
F(\hat U)=[(\theta-\theta_0)\xi_T-\xi\theta_{0T}-\sigma_{0T}]\hat U-i[\hat U_T+\hat U].\label{eq24}
\end{eqnarray}
We  assume  Poincar\'e-type asymptotic expansion for $\hat U$ as $\hat U (\theta,T;\gamma)=\sum_{n=1}^{\infty}\gamma^n \hat U_n (\theta,T)$  and further restrict ourselves to calculation of order $(\gamma)$ such that $\hat U(\theta,T;\gamma)=\hat U_0(\theta,T)+\gamma\hat U_1(\theta,T)$, where $\hat U_0=\eta \mbox{sech}[\eta(\theta-\theta_0)]$. Further, we assume that  $\hat U_1=\phi_1+i\psi_1$, where $\phi_1$ and $\psi_1$ are real. On substituting the above, in  Eqs. (\ref{eq23}) and (\ref{eq24}), we obtain 
\begin{subequations}
\begin{eqnarray}
L_1 \phi_1\equiv -\eta^2 \phi_1+\phi_{1\theta\theta}+6\hat|U_0|^2 \phi_1=\mbox{Re} F(\hat U_0),\label{eq25a}\\
L_2 \psi_1\equiv -\eta^2 \psi_1+\psi_{1\theta\theta}+2\hat|U_0|^2 \psi_1=\mbox{Im} F(\hat U_0),\label{eq25b}
\end{eqnarray}
\end{subequations}
where 
\begin{subequations}
\begin{eqnarray}
\mbox{Re} F(\hat U_0)&=&[(\theta-\theta_0)\xi_T-\xi\theta_{0T}-\sigma_{0T}]\hat U_0,\label{eq26a}\\
\mbox{Im} F(\hat U_0)&=&-[\hat U_{0T}+\hat U_0]. \label{eq26b}
\end{eqnarray}
\end{subequations}
In Eqs. (\ref{eq25a}) and (\ref{eq25b}), $L_1$ and $L_2$ are self-adjoint operators.  It may be checked  that, the solutions of the homogeneous part of Eqs. (\ref{eq25a}) and (\ref{eq25b}) are $\hat U_{0\theta}$ and $\hat U_0$ and hence we have the following secularity conditions.
\begin{subequations}
\begin{eqnarray}
\int_{-\infty}^{\infty} \hat U_{0\theta} \mbox{Re} F(\hat U_0) d\theta =0,\label{eq27a}\\
\int_{-\infty}^{\infty} \hat U_{0} \mbox{Im} F(\hat U_0) d\theta =0.\label{eq27b}
\end{eqnarray}
\end{subequations}
On evaluating the above integrals after substituting the values of $\hat U_{0\theta},\hat U_0, \mbox{Re} F(\hat U_0) $ and $\mbox{Im} F(\hat U_0) $ , we obtain $\xi_T=0,~ \eta_T=-2\eta$, which can be written
in terms of the original time variable $ t$ after integrating once as
\begin{eqnarray}
\xi=\xi_0,~\eta=\eta_0 e^{-2\gamma t},\label{eq28}
\end{eqnarray}
where $\eta_0$ and $\xi_0$ are the initial amplitude and velocity of the soliton. The first of Eq. (\ref{eq28}) says that when the protein-DNA system interacts with  the surrounding viscous medium  the velocity of the soliton remains constant. However, the second of Eq. (\ref{eq28}) says that, the amplitude of the soliton decreases as  soliton propagates. In other words,  the viscous nature of the solvent medium damps out the soliton exponentially and hence the soliton is expected to travel  only a short distance.\\
Now, we construct the perturbed soliton by solving  Eqs. (\ref{eq25a}) and (\ref{eq25b}) . For that,  first we solve the homogeneous part of Eq. (\ref{eq25a}), which admits the following two  particular solutions.
\begin{subequations}
\begin{eqnarray}
\phi_{11}&=&\mbox{sech}\eta(\theta-\theta_0)\tanh\eta(\theta-\theta_0),\label{eq29a}\\
\phi_{12}&=&\frac{1}{\eta}[\frac{3}{2}\eta(\theta-\theta_0)\mbox{sech}\eta(\theta-\theta_0)\tanh\eta(\theta-\theta_0)\nonumber\\
&&+\frac{1}{2}\tanh\eta(\theta-\theta_0)\sinh\eta(\theta-\theta_0)\nonumber\\
&&-\mbox{sech}\eta(\theta-\theta_0)].\label{eq29b}
\end{eqnarray}
\end{subequations}
The general solution can be found out by using the following expression.
\begin{eqnarray}
\phi_1 &=&C_1 \phi_{11}+C_2\phi_{12}-\phi_{11}\int_{-\infty}^{\theta}\phi_{12} \mbox{Re} F(\hat U_0) d\theta\nonumber\\
&&+\phi_{12}\int_{-\infty}^{\theta} \phi_{11} \mbox{Re} F(\hat U_0) d\theta. \label{eq30}
\end{eqnarray}
Here $C_1$ and $C_2$ are arbitrary constants. We construct $\phi_1$ by substituting the values of $\phi_{11}, ~\phi_{12}$ and $\mbox{Re} F(\hat U_0)$ given in Eqs. (\ref{eq29a}), (\ref{eq29b}) and (\ref{eq26a}) in Eq. (\ref{eq30}) and after evaluating the integrals, we get
\begin{eqnarray}
\phi_1 &=&-\frac{1}{\eta}\left[C_2+\frac{1}{2}(\xi \theta_{0T}+\sigma_{0T})\right]\mbox{sech}{\eta(\theta-\theta_0)}+\left[C_1+\frac{3C_{2}}{2}(\theta-\theta_0)+\frac{1}{2}(\theta-\theta_0)\right.\nonumber\\
&&\left(\xi\theta_{0T}+\sigma_{0T})\right]\mbox{sech}\eta(\theta-\theta_0)\tanh\eta(\theta-\theta_0)+\frac{C_2}{2\eta}\sinh\eta(\theta-\theta_0)\tanh\eta(\theta-\theta_0).\label{eq31}
\end{eqnarray}
The last term in Eq.(\ref{eq31}) is a secular term that leads to a solution  which is unbounded and hence, it is removed by choosing the arbitrary constant $C_2=0$. Further, by applying the boundary conditions ${\phi_1}|_{\theta=\theta_0}=$constant = $c$ and ${\phi_{1\theta}} |_{\theta=\theta_0}=0$, we obtain $\frac{1}{\eta}(\xi\theta_{0T}+\sigma_{0T})=-c$ and  $C_1=0$. Using the above results in Eq. (\ref{eq31}), the general solution $\phi_1$ is written as  
\begin{eqnarray}
\phi_1&=& c [1-(\theta-\theta_0)\tanh\eta(\theta-\theta_0)]\mbox{sech}\eta (\theta-\theta_0).\label{eq32}
\end{eqnarray}

Next, we solve  Eq.(\ref{eq25b}),  the homogeneous part of which admits the following particular solutions.
\begin{subequations}
\begin{eqnarray}
\psi_{11}&=&\mbox{sech}\eta(\theta-\theta_0),\label{eq34a}\\
\psi_{12}&=& \frac{1}{2\eta}[\eta(\theta-\theta_0)\mbox{sech}\eta(\theta-\theta_0)+\sinh\eta(\theta-\theta_0)].\label{eq34b}
\end{eqnarray}
\end{subequations}
 Knowing two particular solutions,  the general solution of Eq.(\ref{eq25b}) can be found from
\begin{eqnarray}
\psi_1&=&C_3\psi_{11}+C_4\psi_{12}-\psi_{11}\int_{-\infty}^{\theta}\psi_{12} \mbox{Im} F(\hat U_0) d\theta \nonumber\\
&&+\psi_{12}\int_{-\infty}^{\theta} \psi_{11} \mbox{Im} F(\hat U_0) d\theta, \label{eq35}
\end{eqnarray}
where $C_3$ and $C_4$ are arbitrary constants. We construct the explicit form of $\psi_1$ by substituting the values of $\psi_{11}, \psi_{12}$ and $\mbox{Im} F(\hat U_0)$ given in Eqs. (\ref{eq34a}), (\ref{eq34b}) and (\ref{eq26b}) and evaluating the integrals. 
\begin{eqnarray}
\psi_1 &=&\left\{C_3+\frac{C_4}{2}(\theta-\theta_0) -\frac{\eta}{2} [(\theta-\theta_0)\{\frac{\eta_T}{2\eta}(\theta-\theta_0)-\theta_{0T}\}+\theta_{0T}\tanh\eta(\theta-\theta_0)\right.\nonumber\\
&&\left.+\theta_{0T}(\theta-\theta_0)\mbox{sech}^2\eta(\theta-\theta_0)]\right\}\mbox{sech}(\theta-\theta_0)\nonumber\\
&&+\frac{C_4}{2\eta}\sinh\eta(\theta-\theta_0).\label{eq35a}
\end{eqnarray}
 The above solution for  $\psi_1$ contains  secular term  that is a term proportional to $\sinh\eta(\theta-\theta_0)$  which can be  removed by choosing $C_4=0$. Further, we obtain $C_3=0$ and $\theta_{0T}=0$ upon using the boundary conditions ${\psi_1}|_{\theta=\theta_0}=0$ and ${\psi_{1\theta}} |_{\theta=\theta_0}=0$.  On using the above results in Eq. (\ref{eq35a}), the final form of $\psi_1$ is written as
\begin{eqnarray}
\psi_1=\frac{\eta}{2} (\theta-\theta_0)^2 \mbox{sech}\eta(\theta-\theta_0).\label{eq36}
\end{eqnarray}
 Using the results given in Eqs.(\ref{eq32}) and (\ref{eq36}) we write down the final form of the first order perturbed soliton $U=(\hat U_0+\gamma (\phi_1+i\psi_1)) \exp[i\xi(\theta-\theta_0)+i(\sigma-\sigma_0)]$ (by choosing $\gamma =1$) as
\begin{eqnarray}
U&=&\left[\eta\mbox{sech}\eta(\theta-\theta_0)+c
[1-(\theta-\theta_0)\tanh\eta(\theta-\theta_0)]\right.\nonumber\\
&&\left.+i\frac{\eta}{2}(\theta-\theta_0)^2 \mbox{sech}\eta(\theta-\theta_0)\right] \exp[i\xi(\theta-\theta_0)+i(\sigma-\sigma_0)].\label{eq37}
\end{eqnarray} 
\begin{figure}
\begin{center}
\epsfig{file=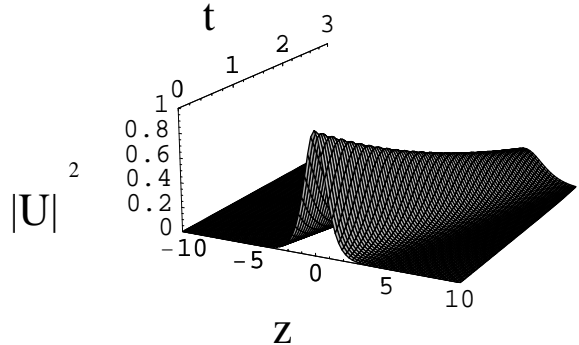,height =5cm, width=8cm}
\caption{ Square of absolute value of the perturbed soliton solution (Eq. (\ref{eq37})) under   viscous damping.}
\end{center}
\end{figure} 
 In Fig. 6, we have plotted the square of the absolute value of the perturbed soliton i.e. $|U|^2$ from 
 Eq.(\ref{eq37}). From the figure,  we observe that the amplitude of the soliton decreases as time progresses, because of the damping due to viscosity of the surrounding medium. Therefore, when the viscosity is high the soliton is expected to travel only for a short time and will stop after that. On the other hand, when the viscosity is low, the soliton will travel for some time. Similar results have also been observed by Yakushevich \cite{ref19} through numerical analysis. They showed that when the viscosity is low, the soliton passes more than 3000 chain links in DNA like a heavy Brownian particle and when the viscosity is large, the soliton stops after a few chain links. 
\section{Conclusion}
In this paper, we investigated the nonlinear dynamics of  a protein-DNA molecular system under thermal fluctuations in a   viscous surrounding medium   by considering  DNA as  a set of  two coupled linear chains and protein as a single linear molecular chain interacting through linear coupling.  In the non-viscous limit, the dynamical equation for the system  is derived from the Hamiltonian through a semiclassical approach using Glauber's coherent state method combined with Holstein-Primakoff (H-P) bosonic representation under continuum approximation. The equation of motion reduces to a set of  coupled  equations in which the equation for  DNA dynamics is a nonlinear equation for rotation of bases and  inhomogeneous   wave equations representing   vibration of hydrogen atom in the bases and  for the   protein molecule. In the linear limit, the above equations are decoupled and    reduced to  time-dependent Schr\"{o}dinger equation  for a free particle and one-dimensional homogeneous linear wave equations.  While the former one admits     dispersive plane transverse wave solution the later ones give non-dispersive wave solutions.    When  protein molecule and thermal phonons started interacting with the DNA,  the  coupling introduces nonlinearity into the dynamics of bases in DNA, and  the set of coupled equations of motion  reduce to the completely integrable nonlinear Schr\"{o}dinger  equation that admits N-soliton solutions.  During interaction the energy of the excited DNA molecule increases and the nonlinearity localizes the energy thus forming localized solitons. The solitons represent opening of base pairs   in both the strands  which collectively  form a bubble travelling along the DNA double helical chain at physiological temperature.   Thus, the protein molecule acts as a zip runner that opens the base pairs which close when the protein molecule progress  along the DNA chain.   For, strong coupling, the  amplitude of the soliton is expected to decrease.  On the other hand, the dynamics of the system in the viscous medium  is governed by a perturbed nonlinear Schr\"{o}dinger equation.  The effect of viscosity is understood by carrying out a multiple scale perturbation analysis. The results show that while the amplitude of the soliton decreases  and the velocity  remains constant as time progresses. The soliton,  damps out  quickly in the case of high viscosity and moves for some time when the viscosity of the surrounding medium is low.   The events that happen in the  present study may represent the binding of an RNA-polymerase to a promoter site in the DNA during the transcription process. Our results have very strong coincidence with the experimental data \cite{ref27,ref28,ref29,ref30,ref31} that the binding of RNA-polymerase to the promoter site in DNA is accompanied by a local distorsion of the DNA bases in the form of solitons which can propagate along the DNA double helix.  In nature, protein binds to DNA in a very specific site  like promoter, coding or  terminator which has a specific sequence of bases and this 
    makes the strands
   site-dependent or inhomogeneous.  Hence, it is  important to understand  the nonlinear
   dynamics of inhomogeneous DNA with  the protein bound  to specific site of DNA  and the study is under progress.
\section*{Acknowledgements}
   The work of M. D and V.V   forms  part of a major  DST  project. 


\begin{thebibliography}{03} 
  \bibitem{ref1}
  L. Stryer, {\it Biochemistry. $4^{th}$ ed} (W. H. Freeman and Company, New York, 1995). 
\bibitem{ref2}
S. J. van Noort, K. O. van der Werf, A. P. M. Eker, C. Wyman, B. G. de Grooth, N. F. van Hulst, and J. Greve, Biophys. J. {\bf 74}, 2840 (1998).
\bibitem{ref3} 
S. J. Koch and M. D. Wang. Phys. Rev. Lett. {\bf 91}, 028103 (2003).
  \bibitem{ref4}
  S. Klimasauskas, S. Kumar, R. J. Roberts and X. Cheng, Cell, {\bf{36}}, 357 (1994).
   \bibitem{ref5}  
  S. Klimasauskas, T. Szyperski, S. Serva and K. Wuthrich, EMBO J. {\bf{17}}, 317 (1998).
   \bibitem{ref6}
K. Liebert, A. Hermann, M. Schlickenrieder and A. Jeltsch, J. Mol. Biol. {\bf{341}}, 443 (2004).
\bibitem{ref6a}
N. Shimamoto, J. Biol. Chem. {\bf 274}, 15293 (1999).
\bibitem{ref7}
N. Huang, N. K. Banavali and A. D. MacKerell Jr, Proc. Natl. Acad. Sci. {\bf{100}}, 68 (2003).
\bibitem{ref8}
J. R. Horton, G. Ratner, N. K. Banavali, N. Huang, Y. Choi, M. A. Maier, V. E. Marquez, A. D. MacKerell Jr
and X. Cheng, Nucleic Acids Res. {\bf{32}}, 3877 (2004).
\bibitem{ref9}
N. Huang and A. D. MacKerell Jr, J. Mol. Biol. {\bf{345}}, 265 (2005).
\bibitem{ref10}
M. Karplus and G. A. Petsko, Nature, {\bf{347}}, 631 (1990).
 \bibitem{ref11}  
S. W. Englander, N. R. Kallenbanch, A. J. Heeger, J. A. Krumhansl and S. Litwin,
  Proc. Natl. Acad. Sci. U.S.A {\bf77}, 7222 (1980).
\bibitem{ref12}
S. Yomosa, Phys. Rev. A {\bf 27}, 2120 (1983).
\bibitem{ref13}
S. Yomosa, Phys. Rev. A {\bf 30}, 474 (1984).
\bibitem{ref14}
S. Takeno and S. Homma,  Prog. Theor. Phys. {\bf 70}, 308 (1983).
\bibitem{ref15}
S. Takeno and S. Homma,  Prog. Theor. Phys. {\bf 72}, 679 (1984). 
\bibitem{ref16}
M. Peyrard and A. R. Bishop,  Phys. Rev. Lett. {\bf 62}, 2755 (1989).
\bibitem{ref17}
P. L. Christiansen, P. C. Lomdahl and V. Muto,  Nonlinearity,  {\bf 4}, 477 (1991). 
\bibitem{ref18}
L. V. Yakushevich,  Nanobiology, {\bf 1}, 343 (1992).
\bibitem{ref20}
   J. A. Gonzalez and M. M. Landrove, Phys. Lett. A {\bf{292}}, 256 (2002). 
  \bibitem{ref21}
  M. Peyrard, Nonlinearity, {\bf{17}}, R1 (2004).   
\bibitem{ref22}
M. Daniel and V. Vasumathi, Physica D, {\bf 231}, 10 (2007).
\bibitem{ref23}
M. Daniel and V. Vasumathi, Phys. Lett. A {\bf 372}, 5144 (2008).
\bibitem{cm}
A. Campa, Phys. Rev. E {\bf 63}, 021901 (2001).
\bibitem{jm1}
S. Cocco, R. Monasson and J. F. Marko, Phys. Rev. E {\bf 65}, 041907 (2002).
\bibitem{jm2}
S. Cocco , R. Monasson and J. F. Marko, Phys. Rev. E {\bf 66}, 051914 (2002).
 \bibitem{ref19}
  L. V. Yakushevich, A. V. Savin and L. I. Manevitch, Phys. Rev . E {\bf{66}}, 016614 (2002).
\bibitem{sm1}
S. Zdravkovic and M. V. Sataric, Phys. Scr. {\bf 64}, 612 (2001).
\bibitem{sm2}
S. Zdravkovic and M. V. Sataric, J. Mod. Phys. B {\bf 17}, 5911 (2003).
\bibitem{vs1}
C. B. Tabi, A. Mohamadov and T. C. Kofane, J. Math. Biosci. Eng. {\bf 5}, 205 (2008).
\bibitem{dvs}
A. S. Davydov, Ukr. Fiz. Zh. {\bf 20}, 179 (1975).
\bibitem{dvs1}
A. S. Davydov, {\it Solitons in Molecular Systems} (Reidel, Dordrecht, 1985).
\bibitem{dvs2}
A. C. Scott, Physica Scripta {\bf 29}, 279 (1984).
\bibitem{dvs3}
J. X. Xiao and L. Yang, Phys. Rev. A {\bf 44}, 8375 (1991).
\bibitem{dvs4}
M. Daniel and K. Deepamala, Physica A {\bf 240}, 526  (1997).
\bibitem{dvs5}
M. Daniel and M. M. Latha, Physica A {\bf 298}, 351 (2001).
\bibitem{toda1}
S. Yomosa, J. Phys. Soc. Jpn. {\bf 53}, 3692 (1984).
\bibitem{toda2}
M. V. Sataric, L. Matsson, and J. A. Tuszynski, Phys. Rev. E {\bf 74 }, 051902 (2006).
\bibitem{sp}
M. D. Betterton and F. Julicher, Phys. Rev. Lett. {\bf 91}, 258103 (2003).
\bibitem{sp1}
J. Yan and J. F. Marko, Phys. Rev. E {\bf 68}, 011905 (2003).
\bibitem{sp2}
S. M. Bhattcharjee, Europhys. Lett. {\bf 65}, 574 (2004).
\bibitem{sp3}
F. Habib and R. Bundschuh, Phys. Rev. E {\bf 72}, 031906 (2005).
\bibitem{muru}
R. Murugan, Phys. Rev. E {\bf 76}, 011901 (2007).
\bibitem{ref24}
M. V. Sataric and J. A. Tuszynski, Phys. Rev. E {\bf 65}, 051901 (2002).
\bibitem{mdlk}
M. Daniel and L. Kavitha, Phys. Lett. A {\bf 295}, 121 (2002) and references therein.
\bibitem{hol1}
N. S. Fialko and V. D. Lakhno, Phys. Lett. A {\bf 278}, 108 (2000).
\bibitem{hol2}
A. M. Guo and H. Xu, Phys. Lett. A {\bf 364}, 48 (2007).
\bibitem{hp}
T. Holstein and H. Primakoff, Phys. Rev. {\bf 58}, 1098 (1940).
\bibitem{glab}
R. J. Glauber, Phys. Rev. {\bf 131}, 2766 (1963).
\bibitem{ref25}
V. E. Zakharov and A. B. Shabat, Sov. Phys. JETP {\bf 34}, 62  (1972).
\bibitem{ref26}  
T. Ha, I. Rasnik, W. Cheng, H.P. Babcock, G. H. Gauss, T. M. Lohman  and S. Chu, Nature, {\bf 419}, 638 (2002). 
\bibitem{ref27}
 U. Siebenlist, R. B. Simpson and W. Gilbert, Cell {\bf 20}, 269 (1980).
 \bibitem{ref28}
 U. Siebenlist, Nature {\bf 279}, 651 (1979).
 \bibitem{ref29}
 M. Sluyser, Trends  Biochem. Sci. {\bf 8}, 236 (1983).
\bibitem{ref30}
 J.W. Saucier and J. C. Wang, Nature. New Biol. {\bf 239}, 167 (1972).
 \bibitem{ref31}
 S. G. Kamzolova, Stud. Biophys. {\bf 87}, 175 (1982). 
\bibitem{mp1}
M. Peyrard, J. Biol. Phys. {\bf 27}, 217 (2001).
\bibitem{mdjb}
M. Daniel and J. Beula, Phys. Rev. B {\bf 77}, 144416 (2008).
\end{thebibliography}
\end{document}